\documentclass{article}
\usepackage{hiph-art}
\usepackage{graphicx}
\usepackage{amssymb}
\volnumber{19} \issuenumber{1} \edyear{2004}                             
\frompage{000} \topage{000}                                              
\recrevdate{1 January 2004}                                              
\def\rightmark{Towards the 3D-Imaging}
\def\leftmark{P.\ Danielewicz {\em et al.}}

\title{Towards the 3D-Imaging of Sources}
\authors{{P.\ Danielewicz$^{1,2,a}$, D.\ A.\ Brown$^3$, M.\ Heffner$^3$, S.\ Pratt$^2$
and R.~Soltz$^3$
\index{Danielewicz, P.} 
\index{Brown, D.\ A.} 
\index{Heffner, M.}
\index{Pratt, S.}
\index{Soltz, R.} %
}\\[2.812mm]
{\normalsize
\hspace*{-8pt}$^1$National Superconducting Cyclotron Laboratory, Michigan State University,\\ East Lansing, MI 48824, USA\\[0.2ex]
\hspace*{-8pt}$^2$Department of Physics and Astronomy, Michigan State University,\\ East Lansing, MI 48824, USA\\[0.2ex]
\hspace*{-8pt}$^3$Lawrence Livermore National Laboratory,\\ Livermore, CA 94551, USA
}}

\abstract{Geometric details of a nuclear reaction zone, at the time of particle emission,
can be restored from low relative-velocity particle-correlations, following imaging.
Some of the source details get erased and are a potential cause of problems in the imaging,
in the form of instabilities.  These can be coped with by following the method of discretized
optimization for the restored sources.  So far it has been possible to produce
1-dimensional emission source images, corresponding to the reactions averaged over all possible
spatial directions.  Currently, efforts are in progress to restore angular details.}



\makeindex
\begin{document}

\maketitle

\section{Introduction}\label{intro}
\begin{figure}[htb]
\begin{center}
\includegraphics[width=.53\linewidth]{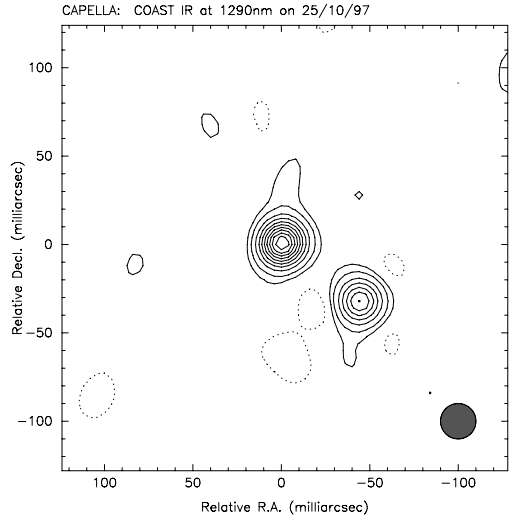}
\end{center}
\vspace*{-.5cm}
\caption[]{Reconstructed image of Capella, from data taken
on October 25, 1997, at wavelength of 1.3 \protect$\mu$m, after
\protect\cite{mon03}.
}
\label{fig1}
\end{figure}

In astronomy, interferometry was initially just used for determining star
diameters or binary-star separations.  Since then, the methodology has moved to
the determination of detailed images of star systems evolving with time, as exemplified in Fig.\
\ref{fig1}.
This concerns the intensity and phase interferometry.
It is needless to say that the
nuclear reaction physics is way behind the astronomy, in extracting
information from particle interference or, generally, correlations.  Besides astronomy,
widely ranging areas, that employ different types of imaging, include police work,
tomography and seismography.

The general task of imaging can be formulated rather simply.  Principally, it amounts to inverting
an integral relation of the form
\begin{equation}
 C(q) = \int dr \, K(q,r) \, S(r) \, .
\end{equation}
Here, $C$ represents the quantity that is measured as a function of $q$ and $S$, which is a function
of $r$, represents an image of interest.  Given the data on $C$, with errors, one needs to determine
$S$, which requires an inversion of the kernel $K$.

\begin{figure}[htb]
\begin{center}
\includegraphics[width=.68\linewidth]{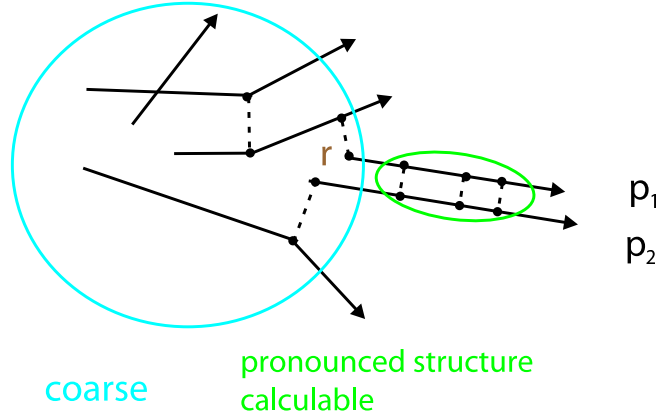}
\end{center}
\vspace*{-.6cm}
\caption[]{The possibility of imaging, in the reactions of heavy nuclei,
results from an interplay of the coarse geometric features of the reaction
zone, with pronounced and calculable geometric features within the two-particle
wave-function.
}
\label{fig2}
\end{figure}

The possibility of imaging in the reactions of heavy nuclei, with many-particle final states,
results from an interplay between the geometric features of the emission zone and of structures
in two-particle wavefunction, usually found at low interparticle velocities, cf.\ Fig.\ \ref{fig2}.
The features in the wavefunction normally change rapidly as the interparticle velocity changes.
If we consider an inclusive cross-section for pair emission, $d\sigma/d{\bf p}_1 \, d{\bf p}_2$,
it is given in terms of a multiparticle amplitude squared, integrated over unobserved particles.
This amplitude can be factored into the outgoing two-particle
wavefunction and a reminder.  The reminder squared and integrated over unobserved particles
represents the source $S'$,
without a strong dependence on the relative
velocity or momentum ${\bf q}$:
\begin{equation}
\frac{d \sigma}{d{\bf p}_1 \, d{\bf p}_2} =  \int  d {\bf r} \,  S'_{\bf P} ({\bf r}) \,
|\Phi^{(-)}_{{\bf q}} ( {\bf r})|^2 \, .
\end{equation}
Here, ${\bf r}$ and ${\bf P}$ are the relative c.m.\ position and total momentum, respectively, and $S'$ may be interpreted as
a distribution of emitted pairs in ${\bf r}$.

If we normalize the two-particle cross section to the product of
single-particle cross sections, we obtain the correlation function:
\begin{equation}
C({\bf q}) =
{{d \sigma \over d{\bf p}_{1} \, d{\bf p}_{2}} \over
{d \sigma \over d{\bf p}_{1}} \, {d \sigma \over d{\bf p}_{2}}} =
\int
d{\bf r} \,
S_{\bf P}({\bf r}) \,
|\Phi_{{\bf q}}^{(-)} ({\bf r})|^2 \, ,
\label{eqC=}
\end{equation}
in terms of the source function $S$ now normalized to 1, $\int d{\bf r}
\, S_{\bf P}({\bf r}) = 1$, since, for large $q$, the correlation function
will approach 1.  For a plane-wave wave-function, we have $|\Phi|^2=1$ and expect $C\simeq 1$.
On the other hand, any nonuniformities in $|\Phi|^2$ can be used as a lens for testing $S$.

The source function is subject to different expectations depending on conditions
in a reaction.  Thus a rapid freeze-out, with not much collective motion developed
in a reaction, should lead to a compact source that is roughly spherical in shape.  A rapid freeze-out, combined
with a strong collective motion, may lead to an oblate shape perpendicular to the velocities
of emitted particles.  If many of the measured particles stem from secondary decays, the source
function will exhibit an extended tail.  If the particles are emitted from a long-lived residual system,
the source function will develop a smutch shape along the particle velocities.  By studying the source
shape, one can learn about the nature of the reaction.

\section{Imaging for the Reactions}\label{imaging}

In practice, information is only contained in the deviation of the correlation function from
unity.  If we subtract 1 from the sides of Eq.\ \ref{eqC=}, we get an imaging relation \cite{bro97}:
\begin{equation}
\mathcal{R}_{\bf P}({\bf q})  =
        C_{\bf P}({\bf q}) -1
=  \int d{\bf r} \left(|\Phi^{(-)}_{\bf {q}}({\bf r})|^2-1\right) \,
        S_{\bf P}({\bf r})
=  \int d{\bf r} \, K({\bf q},{\bf r})
\, S_{\bf P}({\bf r}) \, .
\label{eqR=}
\end{equation}
Source imaging is possible when $|\Phi^{(-)}_{\bf {q}}({\bf r})|^2$ deviates from~1,
as due to symmetrization or interaction within the
pair.

Upon spin averaging, the kernel $K$ depends only on the relative angle between
${\bf q}$ and ${\bf r}$.  This is important in the angular decomposition
of $S$ and $\mathcal{R}$ and in the decomposition of the
imaging relation.  Upon decomposing the kernel and the functions according to
\begin{equation}
K({\bf q},{\bf r}) =
\sum_\lambda (2 \lambda +1 )\, K_\lambda (q,r) \, P^\lambda
(\cos{\theta}) \, ,
\end{equation}
and
\begin{equation}
{\mathcal R} ({\bf q}) = \sqrt{4 \pi} \sum_{\lambda
m} {\mathcal R}^{\lambda m} (q) \, {\rm Y}^{\lambda m} (\hat{\bf q}) \, , \hspace*{1.1em}
S({\bf r}) = \sqrt{4 \pi} \sum_{\lambda
m} S^{\lambda m} (r) \, {\rm Y}^{\lambda m} (\hat{\bf r}) \, ,
\end{equation}
we get, from the relation (\ref{eqR=}),
\begin{equation}
{\mathcal R}_{\bf P}^{\lambda m}(q)
= 4 \pi \int dr \, r^2 \, K_\lambda (q, r) \,
S_{\bf P}^{\lambda m} (r) \, .
\label{eqRl=}
\end{equation}
The above shows that different multipolarities of deformation for the source and correlation functions
are directly related to each other.  The $\lambda=0$ version of the relation is
\begin{equation}
{\cal R}_{P}({q})  =  4 \pi
\int dr \, r^2 \,
         K_0 ({q},{r}) \,
S_P(r) \, ,
\end{equation}
where ${\mathcal R}(q)$, $K_0$ and $S(r)$ are the angle-averaged correlation, kernel and source, respectively.

\begin{figure}[htb]
\begin{center}
\vspace*{.1in}
\includegraphics[width=.58\linewidth]{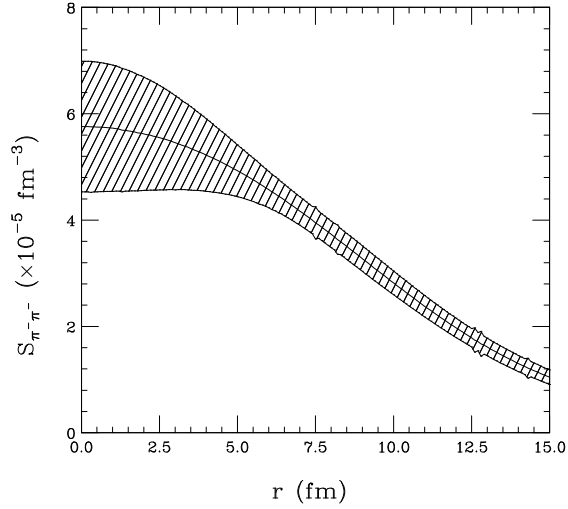}
\end{center}
\vspace*{-.55cm}
\caption[]{Relative source of negative pions in Au + Au collisions at 10.8 GeV/nucleon,
from Fourier transformation of the negative-pion correlation function obtained by
the E877 Collaboration \protect\cite{bar97}.
}
\label{fig3}
\end{figure}

The simplest case of imaging corresponds to a situation where the structure of $|\Phi|^2$ is due
to a pure interference, e.g.\ neutral pions or gammas.  For a symmetrized wavefunction,
\begin{equation}
\Phi^{(-)}_{\bf q}({\bf r})=\frac{1}{\sqrt{2}}\left( e^{i {\bf q} \cdot {\bf r}}
        +e^{-i {\bf q}\cdot {\bf r}}\right)
\end{equation}
the kernel is
\begin{equation}
K({\bf q},{\bf r}) = \left|\Phi^{(-)}_{\bf q}({\bf r})\right|^2-1 =  \cos{(2 {\bf q} \cdot {\bf r})} \, ,
\label{eqKdcmp}
\end{equation}
and the imaging amounts then to a Fourier-cosine transform
\begin{equation}
{\mathcal R}_{\bf P} ({\bf q}) = \int d{\bf r} \cos{\left(2 {\bf q}
 {\bf r} \right) } \, S_{\bf P} ({\bf r}) \, \Rightarrow \,
S_{\bf P} ({\bf r}) = {1 \over \pi^3} \int d {\bf q} \,
\cos{\left(2{\bf
q} {\bf r}\right)} \,  {\mathcal R}_{\bf P}({\bf q}) \, .
\label{eqRSdcmp}
\end{equation}
The angle-averaged relation also represents a Fourier transformation, since
\begin{equation}
K_0(q,r) =
\frac{\sin{(2 q r)}}{2 q r} \, ,
\end{equation}
and, in general, the relations for different multipolarities represent Fourier-Bessel
transforms, since $K_\lambda(q,r)= (-1)^{\lambda/2}\, j_\lambda (2 q r)$.
As an example of the application of Fourier transformation in imaging, Fig.\ \ref{fig3}
shows the relative source of negative pions obtained from the Coulomb corrected correlation
function from measurements of the E877 Collaboration \protect\cite{bar97}.

For many particle pairs, as e.g.\ $pp$, the interaction can
be neither ignored nor simply corrected for, and the
straightforward Fourier transformation cannot be used.
In the general case of correlation, though, the source discretization with a $\chi^2$ fitting may work.
For this, e.g.\ in the angle-averaged relation, one discretizes the integral to get
\begin{equation}
{\mathcal R}_i = \sum_j 4\pi \, \Delta r
\,
              r_j^2 \, K_0 (q_i, r_j) \, S(r_j)
\equiv        \sum_j K_{ij} \, S_j  \, ,
\end{equation}
and, subsequently, varies the $S_j$ values to minimize
\begin{equation}
\chi^2 = \sum_i  \frac{( \sum_j K_{ij} \, S_j-
              {\mathcal R}_i^{exp}))^2}{\sigma_i^2 } \, .
\end{equation}
The derivative of $\chi^2$ with respect to $S_j$ gives a set of algebraic
eqs.\ for~$S$:
\begin{equation}
 \sum_{ij} {1 \over \sigma_i^2} (K_{ij} \, S_j - {\mathcal
R}_i^{exp}) \, K_{ij} = 0 \, ,
\end{equation}
with the solution, in a matrix form,
\begin{equation}
S = (K^\top K)^{-1} \, K^\top \, {\mathcal R}^{exp} \, .
\end{equation}

\begin{figure}[htb]
\parbox{.5\textwidth}{
\includegraphics[width=.98\linewidth,height=1.1\linewidth]{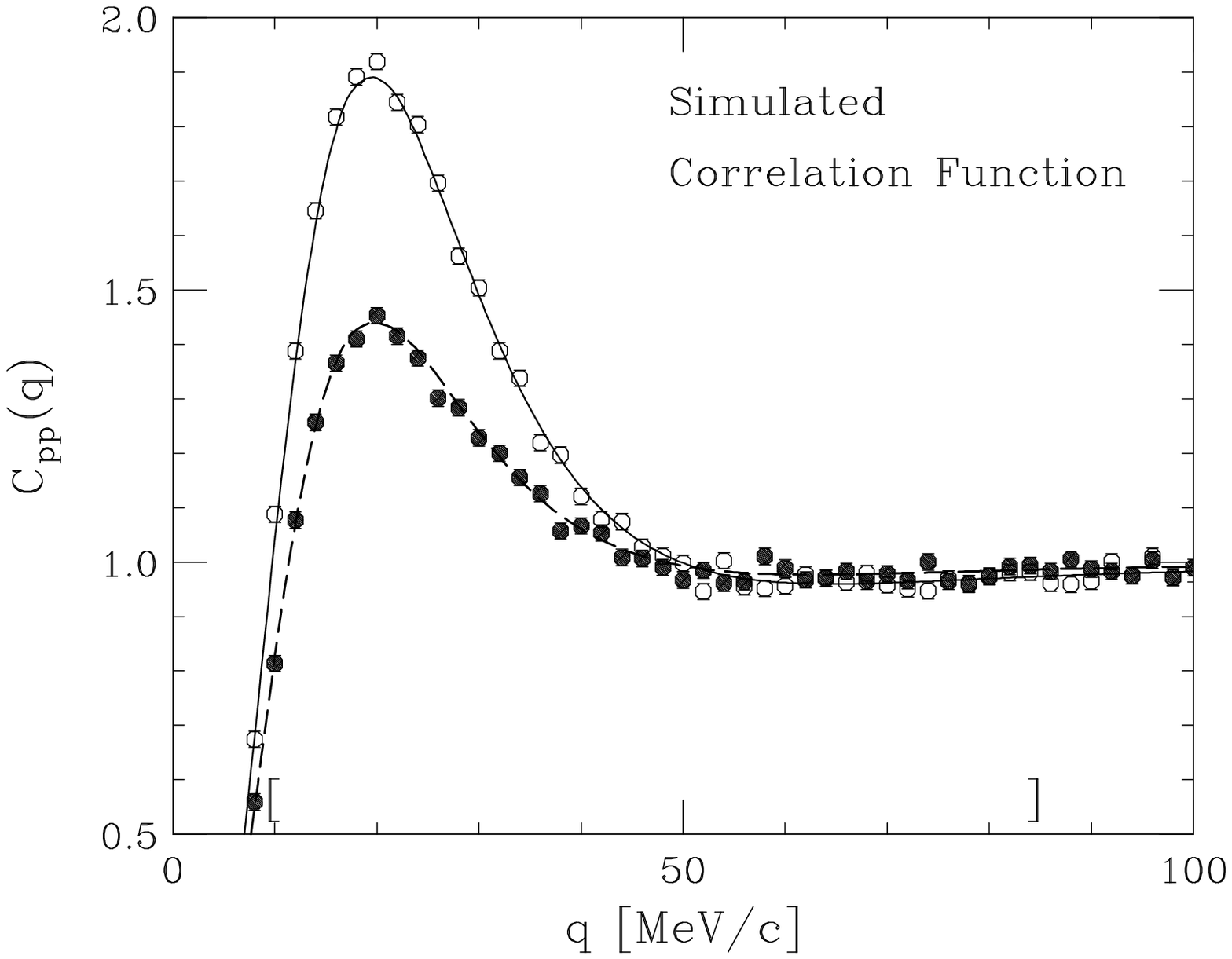}
\hfill
}
\parbox{.5\textwidth}{
\hfill
\includegraphics[width=.98\linewidth,height=.75\linewidth]{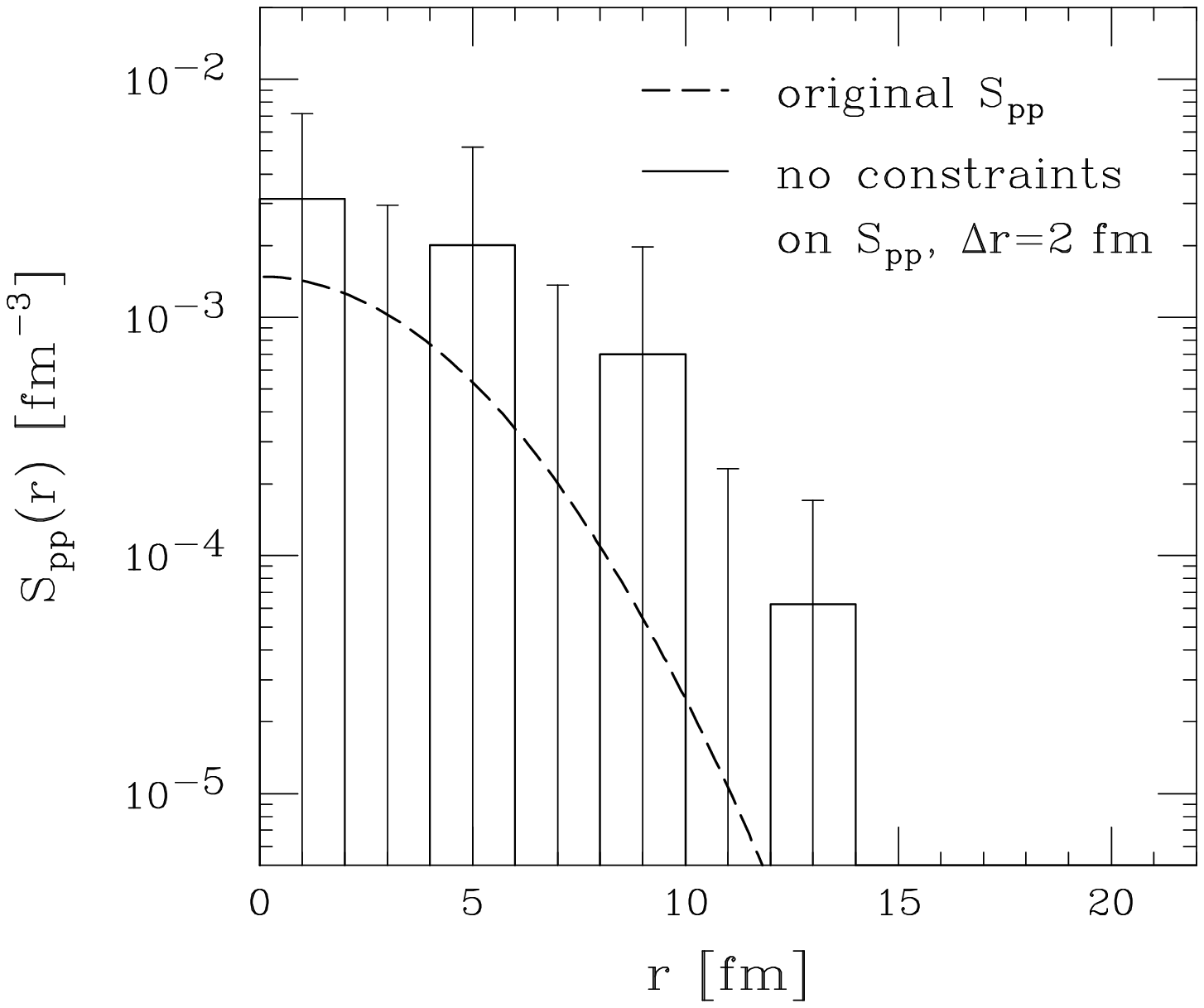}
\hspace*{1em}
\\
\vspace*{1ex}
\hfill
\includegraphics[width=.98\linewidth,height=.75\linewidth]{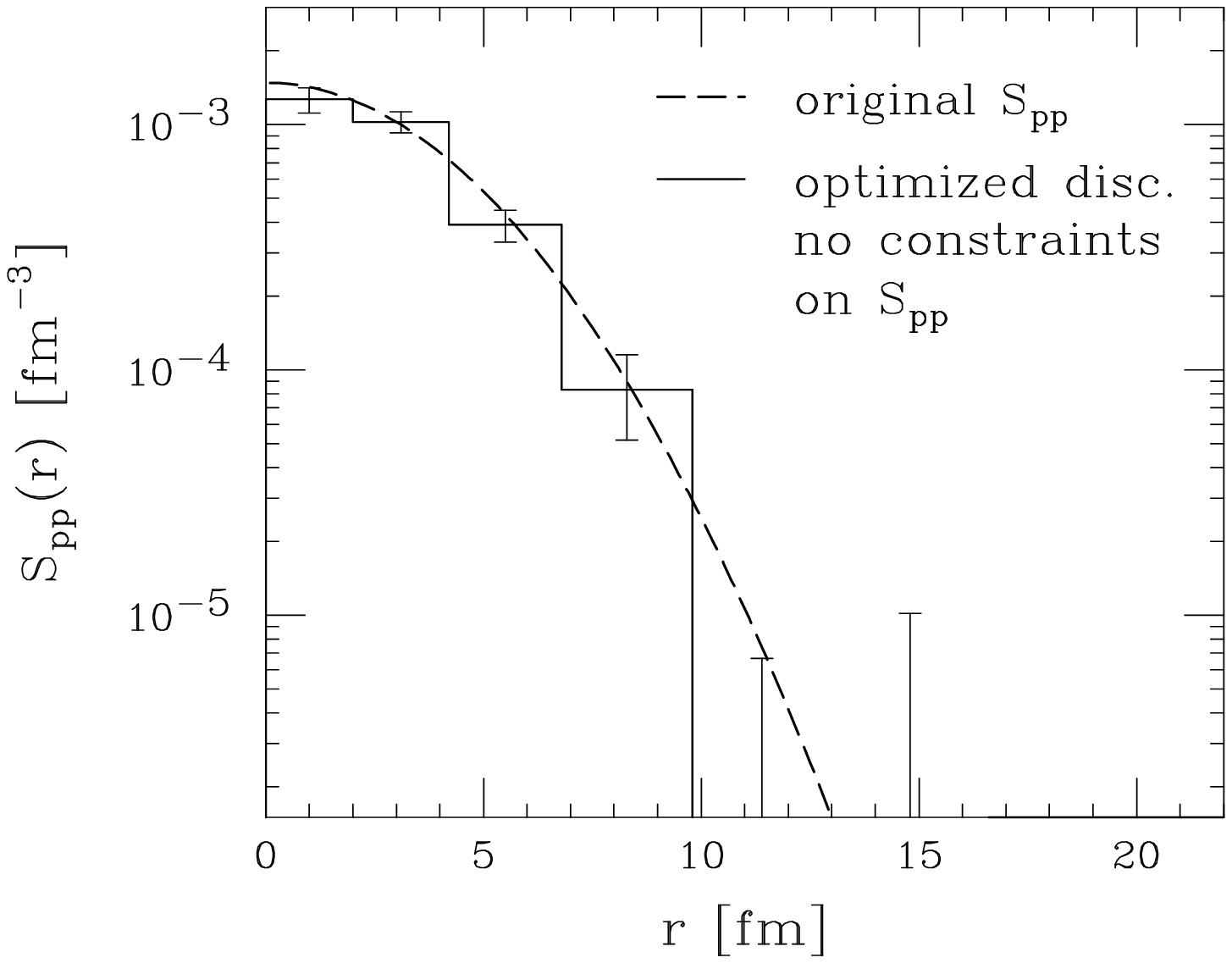}
\hspace*{1em}
\\
}
\vspace*{-1.4cm}
\caption[]{On the left, pp correlation functions constructed from two assumed source
functions, without (lines) and with (symbols) errors added.  On the right, the results
of source restoration through imaging,
from the sharper of the two simulated correlation functions
on the left, following either the discretization with an assumed bin size of 2 fm (right top
figure) or the optimized discretization method \protect\cite{bro97} (right bottom).
}
\label{fig4}
\end{figure}

Figure \ref{fig4} shows the results of simple tests concerning imaging,
where a correlation function is constructed out of an assumed source,
errors are added and inversions are carried out for the source.
This figure illustrates
the problems that can be generally encountered in imaging.  Specifically, the correlation
function may not reflect fine details of the source.  If excessive detail
is demanded within some region, the inversion of the kernel begins to remind
the inversion of a zero, which is signalled by large changes in the output
for small changes in the input.  This situation is encountered in the naive
imaging in Fig.\ \ref{fig4}, relying on a fixed 2 fm bin discretization of
the seeked source.  The situation may be remedied by following the method of
optimized discretization, where the source discretization is varied prior
to the inversion, to ensure minimal anticipated errors.  An exemplary result
of the procedure, together with the original source, is also further shown in Fig.\
\ref{fig4}.

\begin{figure}[htb]
\includegraphics[width=.49\linewidth,height=.48\linewidth]{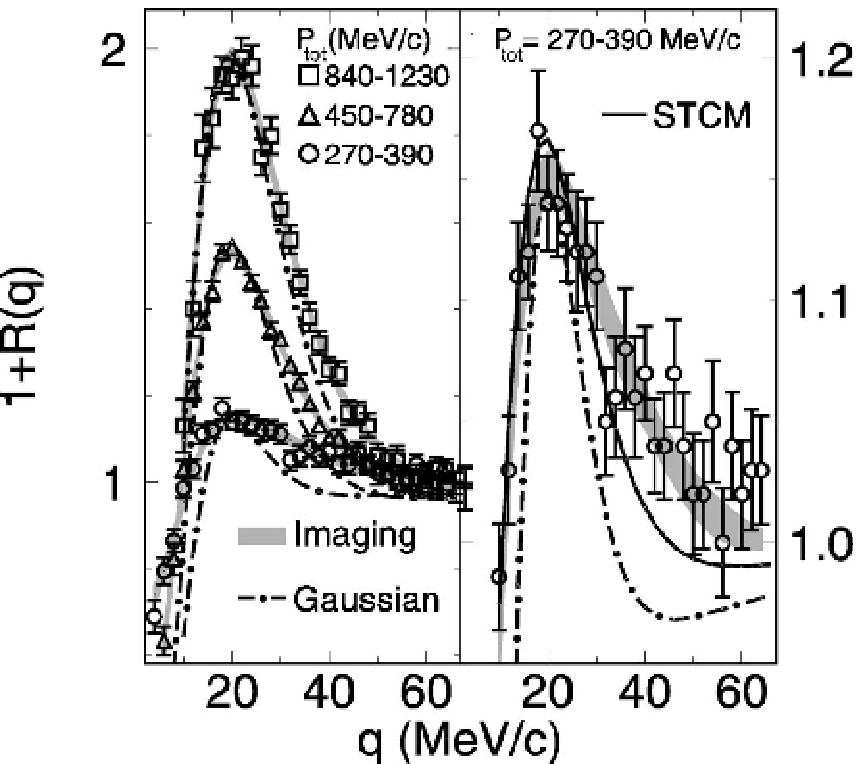}
\hfill
\includegraphics[width=.49\linewidth,height=.55\linewidth]{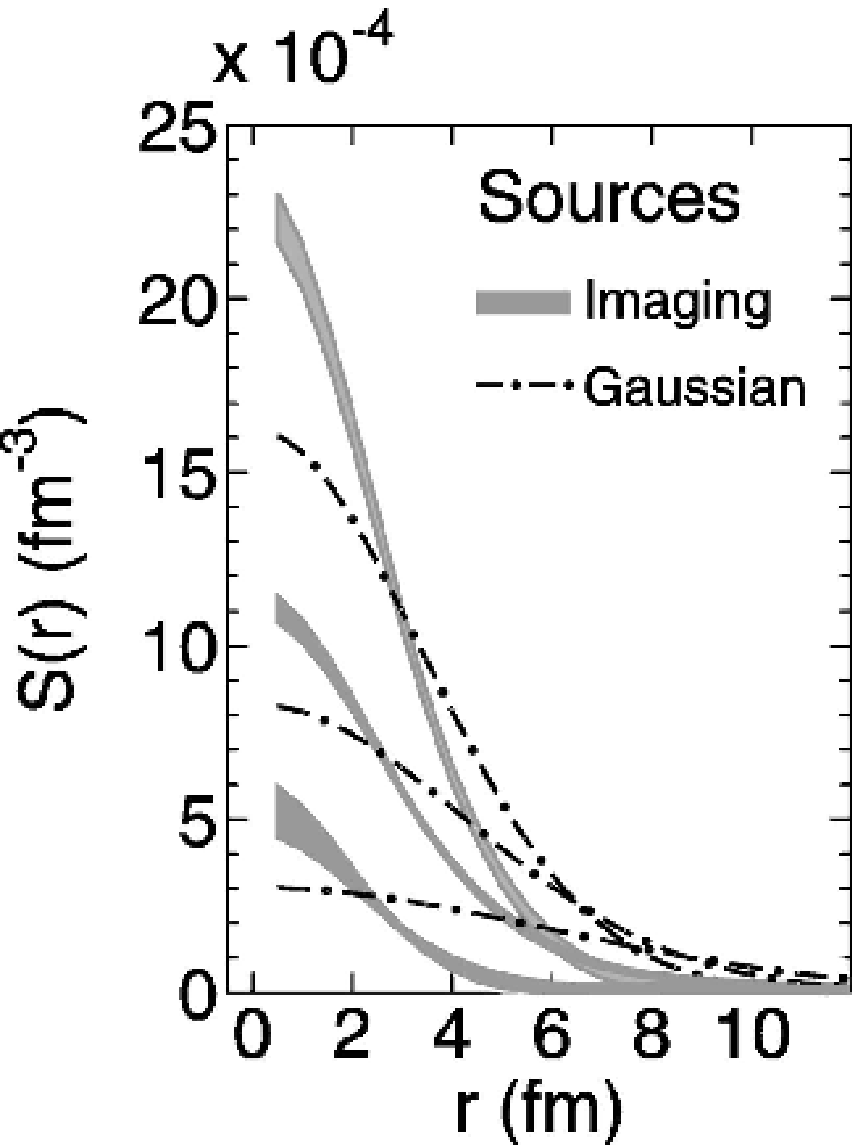}
\vspace*{-.3cm}
\caption[]{On the left, pp correlation functions measured for different
indicated total momentum ranges, in the $^{14}$N + $^{197}$Au reaction at
75 MeV/nucleon.  On the right, the sources imaged (shaded regions) from the correlation
functions on the left, together with the results (lines) when assuming sources in the Gaussian
form, after \cite{ver02}.
}
\label{fig5}
\end{figure}

The imaging has significantly changed the interpretation of different pp correlation data
previously described in terms of Gaussian sources \cite{ver02}, cf.\ Fig.\ \ref{fig5}.
Thus, the short-range portion of sources in the reaction is, generally, non-Gaussian
and the sources exhibit long-range tails.  The weight of the preequilibrium
short-range portion of a source, relative to the long-range halo, changes strongly with
the momentum of an emitted pair, while the spatial extent of the preequilibrium portion
remains relatively constant with the momentum.  The changes in the relative weight of preequilibrium
particles get reflected in the height of the measured correlation function.
The changes in the height of
the correlation function get, though, misinterpreted, in terms of the changes in extent of
the preequilibrium shape, when forcing the Gaussian shape onto a source.

\section{Towards 3D Imaging}

Anisotropies in correlation functions have been used to access differences
in typical emission times for different particles \cite{ghe03} and to access asymmetries in the emission
zone from reaction geometry and collective motion \cite{adl01}.  The basic relation that
can be used to access the anisotropies of the source within imaging is
Eq.\ (\ref{eqRl=}).  For identical particles, only even $\lambda$ contribute to
the angular decompositions in (\ref{eqRSdcmp}) and (\ref{eqKdcmp}).

For the decomposition, we need to choose a reference coordinate system where asymmetries
are assessed and a convenient
system is one with the $z$-axis along the beam axis and the $x$-axis along the net transverse
momentum of the emitted pair.  Either the source $S$ or the correlation $C$ (or
$\mathcal{R}$) may be represented in that system as
\begin{eqnarray}
\nonumber
C({\bf q}) & \hspace*{-.5em} = \hspace*{-.5em} & \sum_{\lambda m} \int d\Omega' \, C({\bf q}') \,
Y_{\lambda m}^*(\hat{\bf q}') \, Y_{\lambda m}(\hat{\bf q})
 =  \sum_{\lambda} \int d\Omega' \, C({\bf q}') \, \mbox{Re} \left(
Y_{\lambda 0}(\hat{\bf q}') \right)
\\ \nonumber
&& \times \,
 \mbox{Re} \left(Y_{\lambda 0}(\hat{\bf q}) \right) +2
\sum_{\lambda, m \ge 1} \int d\Omega' \, C({\bf q}') \, \mbox{Re} \left(
Y_{\lambda m}(\hat{\bf q}') \right) \, \mbox{Re} \left(Y_{\lambda m}(\hat{\bf q}) \right)\\ \nonumber
& \hspace*{-.5em} = \hspace*{-.5em} &  \sum_\lambda \sum_{k_1 \ldots k_\lambda}
T_{k_1 \ldots k_\lambda}^{(\lambda)}(q) \, \hat{\bf q}_{k_1} \ldots \hat{\bf q}_{k_\lambda}
= \int \frac{d \Omega'}{4 \pi} \, C({\bf q}')
+ 3 \int \frac{d \Omega'}{4 \pi} \, C({\bf q}') \, \hat{\bf q}'_k \, \hat{\bf q}_k \\
&& + \frac{15}{2} \int \frac{d \Omega'}{4 \pi} \, C({\bf q}') \, \left( \hat{\bf q}'_k \, \hat{\bf q}'_n
- \frac{1}{3} \, \delta_{kn} \right) \, \hat{\bf q}_k \, \hat{\bf q}_n + \ldots
 \, ,
 \label{eqct}
\end{eqnarray}
where the symmetry with respect to the $y \rightarrow -y$ reflection and the real nature of $C$
have been exploited after the second equality.  (The additional symmetry in a symmetric system at midrapidity
additionally limits the contributions
to only terms characterized by even $\lambda + m$.)  The expression after the third equality is in terms
of cartesian components for spherical tensors of rank $\lambda$.  Finally, the last
expression makes the expansion in terms of tensors of increasing rank explicit.

The leading term on the r.h.s.\ of (\ref{eqct}) is the correlation function averaged over
angles, $C^{(0)}(q)$.  The dipole tensor in the second term on the r.h.s.\ of (\ref{eqct}), i.e.\ a vector, may be expressed as
\begin{equation}
T_k^{(1)}(q) = 3 \int \frac{d \Omega'}{4 \pi} \, C({\bf q}') \, \hat{q}'_k = C^{(1)}(q) \, \hat{\bf n}_k^{(1)}(q) \, ;
\end{equation}
the dipole deformation is
characterized by a direction angle $\theta^{(1)}$ and deformation parameter $C^{(1)}$, both dependent on~$q$.
The quadrupole tensor in the third term can be next expressed as
\begin{eqnarray}
\nonumber
T_{kn}^{(2)}(q) & = & \frac{15}{2} \int \frac{d \Omega'}{4 \pi} \, C({\bf q}') \, \left( \hat{q}'_k \, \hat{q}'_n
- \frac{1}{3} \, \delta_{kn} \right) \\
&=& C_3^{(2)} \, \overline{\overline{{\bf n}_3 \, {\bf n}_3}}^{(2)} + C_1^{(2)} \, \overline{\overline{{\bf n}_1 \, {\bf n}_1}}^{(2)}
-  (C_1^{(2)} + C_3^{(2)}) \, \overline{\overline{{\bf n}_2 \, {\bf n}_2}}^{(2)} \, .
\end{eqnarray}
The quadrupole deformation may be characterized in terms of two deformation parameters,
such as the larger $C_3^{(2)}$ and lower $C_1^{(2)}$, within
the ${\bf P}^\perp$-beam plane, and in terms of the direction angle for the larger deformation, $\theta^{(2)}$.
The axis 2 for the deformation tensor, by symmetry, is perpendicular to the ${\bf P}^\perp$-beam plane.
An analogous decomposition holds for the source.  Overall, for deformations in terms of the lowest angular moments,
there are 6 parameters, as a function of $q$ or $r$, needed in describing the correlations or sources for nonidentical particles,
and 4 parameters for identical particles, since the dipole deformations then vanish.

In terms of numerical analysis, for $\lambda \ge 1$, the same methods can be employed as for $\lambda=0$, including
basis splines and the optimized discretization.  The novel issue encountered for $\lambda \ge 1$ is the theoretical deterioration of
angular resolution as $q \rightarrow 0$ or $r \rightarrow \infty$, amplified by the possible experimental deterioration of resolution
for any practical data.

Figure 6 shows results from a test similar to that in Fig.\ 4, where now a 3-dimensional correlation function has been constructed
from an assumed source for $\pi^-$ pairs \cite{bro04}.  The function
has been modified by errors, an angular decomposition was applied, and then inversion was
employed for each multipolarity and the source was reconstructed.
The original and reconstructed sources are compared along three
orthogonal directions in Fig.\ 6 and it is seen that the imaging is successful at a semiquantitative level.
\begin{figure}[htb]
\centerline{\includegraphics[width=.65\linewidth]{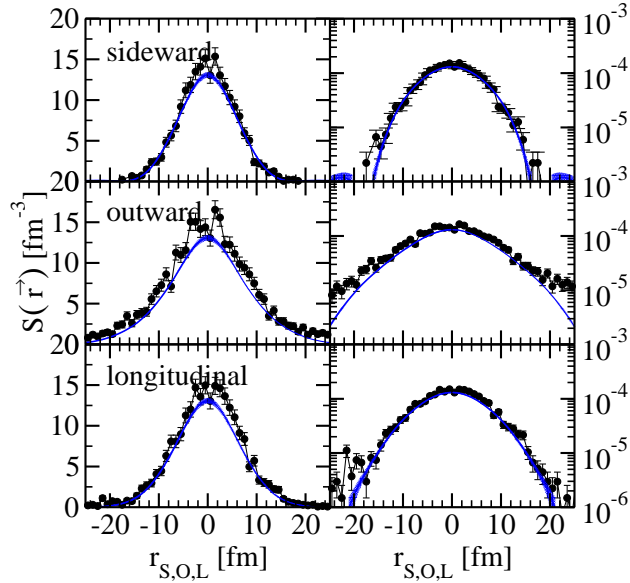}}
\vspace*{-.3cm}
\caption[]{Comparison of the imaged (bands)
and input (symbols with errors representing finite event statistics)
$\pi^--\pi^-$ sources along three perpendicular directions, after~\protect\cite{bro04}.
}
\label{fig6}
\end{figure}

\section{Conclusions}

We have demonstrated the feasibility of imaging the emission sources in heavy-ion reactions.  The one-dimensional imaging
has been carried out for data on pion, proton, kaon and IMF correlations.  The three-dimensional imaging is under practical
developments and we have here described its essential elements.  The methods that are advanced, in particular involving
optimized discretization, permit to investigate source images, on a logarithmic scale, for large separations.
The imaging gives access to the emission source structure.
In their details, the images can contain information pertaining to the phase-space density of particles at freeze-out, produced
entropy, spatial freeze-out density, reaction geometry, developed collective motion, resonance decays and prolonged emission including
phase-transition effects.

\section*{Acknowledgments}
This research is supported by the U.S.\ National Science Foundation under the Grant
PHY-0245009 and by the U.S.\ Department of Energy under the Grant DE-FG02-03ER41259.
Part of this work was performed under the auspices of the U.S.\ Department of Energy
by Lawrence Livermore National Laboratory under Contract W-7405-Eng-48.

\begin{notes}
\item[a]
E-mail: danielewicz@nscl.msu.edu
\end{notes}

\vfill\eject
\end{document}